\documentclass[11pt,superscriptaddress,aps,prd,preprint]{revtex4}
\usepackage{amsmath}

\makeatletter
\usepackage[T1]{fontenc}
\usepackage{amsmath}
\usepackage{amssymb}
\usepackage{graphicx}
\usepackage{tabularx}

\usepackage{slashed}
\usepackage{xcolor}

\newcommand{\bea}{\begin{eqnarray}}
\newcommand{\eea}{\end{eqnarray}}

\begin{document}

%\title{On Lorentz violation in Gravitational Bhabha scattering}
\title{ Lorentz Violation, Gravitoelectromagnetic Field and Bhabha Scattering}

\author{A. F. Santos}\email[]{alesandroferreira@fisica.ufmt.br}
\affiliation{Instituto de F\'{\i}sica, Universidade Federal de Mato Grosso,\\
78060-900, Cuiab\'{a}, Mato Grosso, Brazil}

\author{Faqir C. Khanna\footnote{Professor Emeritus - Physics Department, Theoretical Physics Institute, University of Alberta\\
Edmonton, Alberta, Canada}}\email[]{khannaf@uvic.ca}
\affiliation{Department of Physics and Astronomy, University of Victoria,\\
3800 Finnerty Road Victoria, BC, Canada}

\begin{abstract}

Lorentz symmetry is a fundamental symmetry in the Standard Model (SM) and in General Relativity (GR). This symmetry holds true for all models at low energies. However at energies near the Planck scale, it is conjectured that there may be a very small violation of Lorentz symmetry. The Standard Model Extension (SME) is a quantum field theory that includes a systematic description of Lorentz symmetry violations in all sectors of particle physics and gravity. In this paper SME is considered to study the physical process of Bhabha Scattering in the Gravitoelectromagnetism (GEM) theory. GEM is an important formalism that is valid in a suitable approximation of general relativity. A new non-minimal coupling term that violates Lorentz symmetry is used in this paper. Differential cross section for gravitational Bhabha scattering is calculated. The Lorentz violation contributions to this GEM scattering cross section are small and are similar in magnitude to the case of the electromagnetic field.

\end{abstract}

\maketitle

\section{Introduction}

A remarkably successful description of nature is provided by the Standard Model (SM) and General Relativity (GR). The GR describes the gravitational force at the classical level, while the SM encompasses all others forces of nature up to the quantum level. A fundamental theory that unifies GR and SM is expected to emerge at the Planck scale ($\sim 10^{19}\mathrm{GeV}$). One of the promising candidate for such a theory is the string theory. At this energy scale, minuscule Lorentz-violating effects arise offering a signal for some new physics at the Planck scale \cite{Kost1989,Kost1989_1, Kost1989_2, Kost1989_3, Kost1991, Kost1991_1}. The outstanding problem is the lack of experimental guidance, since direct experiments at the Planck scale are not possible. Due to this difficulty an effective field theory is an appropriate tool to observe any signals of Lorentz violation \cite{Kost1995}. To study such effects an effective field theory, Standard Model Extension (SME), has been constructed \cite{SME1, SME2}. The SME contains the SM, GR (GEM in the present case) and all possible operators that break Lorentz symmetry. The SME is divided into two parts: (i) a minimal extension which has operators with dimensions $d\leq 4$ and (ii) a non-minimal version associated with operators of higher dimensions. Tests of Lorentz invariance at high-energy has been discussed \cite{Coleman}. A complete description of the GR in the framework of the SME, in the context of a Riemann-Cartan spacetime with curvature and torsion, has been studied \cite{Kos0}. This allows us to consider all Lorentz-violating terms involving gravitons coupled to the SM fields.

In the minimal version of SME the electromagnetic sector is composed of two parts: 4 CPT-odd coefficients and 19 CPT-even coefficients. In the gravitational sector there are 19 coefficients for Lorentz violation apart from an unobservable scalar parameter. Such similarity suggests a correspondence between the two sectors, gravitational (GEM) and CPT-even electromagnetic sectors. The existence of an analogy between the electromagnetic sector and the gravitational sector for SME has been developed \cite{QG}. The Lorentz-violating effects modify electromagnetic systems in a similar way as gravitational systems are modified in the weak field approximation. Here this analogy is extended for a non-minimal version of SME.

The search for analogies between electromagnetism and gravity has a long history. For Lorentz invariant theories, the first attempt to unify gravity and electromagnetism started with Faraday in 1832  \cite{Faraday}. Then Maxwell used the similarity between Newton's law and Coulomb's law to formulate a theory of gravitation \cite{Maxwell}. Several others developments have been realized in a similar manner \cite{Heaviside1, Heaviside2, Weyl, Weyl_1, KK, KK_1, Thirring, Matte, Campbell1, Campbell2, Campbell3, Campbell4, Braginsky}. A formal analogy between the gravitational and the electromagnetic fields led to the notion of Gravitoelectromagnetism (GEM) to describe gravitation. The motivations to such theory are: there is a gravitomagnetic field connected with moving masses and the gravitational field speed of propagation is equal to the speed of light.  Therefore, GEM is a formal analogy between Maxwell's equations and a linear approximation, valid under certain conditions, to the Einstein field equations for GR . Many aspects of the usual GEM based on the linearized GR have been discussed \cite{Mashhon}. Various experiments, such as LAGEOS, LAGEOS II, LAGEOS III and the Gravity Probe B have tried to measure the GEM effects, particularly the gravitomagnetic effect\cite{Large, Probe}.

There are three different ways to construct GEM theory: (i) using the similarity between the linearized Einstein and Maxwell equations  \cite{Mashhon}; (ii) based on an approach using tidal tensors \cite{Filipe} and (iii) the decomposition of the Weyl tensor into gravitomagnetic (${\cal B}_{ij}=\frac{1}{2}\epsilon_{ikl}C^{kl}_{0j}$) and gravitoelectric (${\cal E}_{ij}=-C_{0i0j}$) components \cite{Maartens}. Weyl approach is considered. The Weyl tensor is connected with the curvature tensor. It is the traceless part of the Riemann tensor, and it is defined as
\bea
C_{\alpha\sigma\mu\nu}&=&R_{\alpha\sigma\mu\nu}-\frac{1}{2}\left(R_{\nu\alpha}g_{\mu\sigma}+R_{\mu\sigma}g_{\nu\alpha}-R_{\nu\sigma}g_{\mu\alpha}-R_{\mu\alpha}g_{\nu\sigma}\right)\nonumber\\
&+&\frac{1}{6}R\left(g_{\nu\alpha}g_{\mu\sigma}-g_{\nu\sigma}g_{\mu\alpha}\right),
\eea
where $R_{\alpha\sigma\mu\nu}$ is the Riemann tensor, $R_{\mu\nu}$ is the Ricci tensor and $R$ is the Ricci scalar. The analogy between electromagnetism and General Relativity is based on the correspondence $C_{\alpha\sigma\mu\nu} \leftrightarrow  F_{\alpha\sigma}$, where the Weyl tensor is the free gravitational field and $F_{\alpha\sigma}$ is the electromagnetic tensor. The Weyl tensor gives the contribution due to nonlocal sources. The Weyl tensor, together with the energy-momentum tensor, determines tidal accelerations due to the gravitational field,  i.e., Newtonian tidal forces are represented in the Weyl tensor. Combining Einstein equations and Bianchi identities we obtain \cite{Khanna}
\bea
C_{\alpha\beta\mu\nu;}\,^\nu=4\pi G\left(-T_{\mu\beta;\alpha}+T_{\mu\alpha;\beta}+\frac{1}{3}T_{,\alpha}g_{\mu\beta}-\frac{1}{3}T_{,\beta}g_{\mu\alpha}\right),
\eea
where $T_{\mu\nu}$ is the energy-momentum tensor and $T$ is the trace of the energy-momentum tensor. Here, the comma is used to denote the partial derivative and semicolon denotes the covariant derivative.  The Weyl tensor can be decomposed into two symmetric  second rank tensor, 
\bea
{\cal B}_{ij}=\frac{1}{2}\epsilon_{ikl}C^{kl}_{0j}, \quad\quad\quad {\cal E}_{ij}=-C_{0i0j},
\eea
the gravitomagnetic and gravitoelectric field, respectively. These components, in a nearly flat space-time, lead to the Maxwell-like equations, i.e., the GEM equations. The GEM analogy is limited by the fact that the Maxwell field propagates on a given space-time, whereas the gravitational field itself generates the space-time. The other limitation is the fact that the GEM degrees of freedom only depend on the Weyl tensor. 

In the Weyl tensor approach a Lagrangian formulation for GEM has been developed \cite{Khanna}. In this formalism a symmetric gravitoelectromagnetic tensor potential, $A_{\mu\nu}$, which describes the gravitational interaction, is defined. GEM allows scattering processes with gravitons as an intermediate state, in addition to the photon, for electromagnetic scattering. This paper is devoted to the study of gravitational Bhabha scattering in the SME framework.

The Bhabha scattering ($e^+e^-\longrightarrow e^+e^-$) is a process usually used in tests of experiments at high energy accelerators \cite{Venus, Topaz, AMY, Derr, Ari}. To include the Lorentz-violating effects the interaction vertex between fermions and gravitons is modified. A new non-minimal coupling term will be added to the covariant derivative. The Bhabha scattering in the context of CPT-odd non-minimal coupling term has been analyzed \cite{SP, Brito}. Some other applications involving the CPT-odd non-minimal coupling have been developed \cite{Belich1, Belich2, Belich3, Belich4, Belich5}. The CPT-even non-minimal coupling  has been studied in different contexts, such as: cross section of the electron-positron scattering \cite{Casana1}, Dirac equation in the non-relativistic regime \cite{Casana2}, radiative generation of the CPT-even gauge term \cite{Casana3}, and effects induced on the magnetic and electric dipole moments \cite{Casana4}. From the ideas discussed in \cite{QG}, about Lorentz-violating gravitoelectromagnetism, our objective is to calculate the differential cross section for the gravitational Bhabha scattering in the presence of a new Lorentz-violating non-minimal coupling. 

This paper is organized as follows. In section II, a brief introduction to the Lagrangian formalism for  GEM is presented. In section III, the Lorentz-violating GEM is considered. The new interaction vertex is analyzed. In section IV, the differential cross section for gravitational Bhabha scattering with Lorentz violation is determined. In section V, some concluding remarks are presented.

\section{Lagrangian formalism for the GEM}

The Maxwell-like equations of GEM are given as
\bea
&&\partial^i{\cal E}^{ij}=-4\pi G\rho^j,\label{01}\\
&&\partial^i{\cal B}^{ij}=0,\label{02}\\
&&\epsilon^{( ikl}\partial^k{\cal B}^{lj)}+\frac{\partial{\cal E}^{ij}}{\partial t}=-4\pi G J^{ij},\label{03}\\
&&\epsilon^{( ikl}\partial^k{\cal E}^{lj)}+\frac{\partial{\cal B}^{ij}}{\partial t}=0,\label{04}
\eea
where ${\cal E}_{ij}$ is the gravitoelectric field, ${\cal B}_{ij}$ is the gravitomagnetic field, $G$ is the gravitational constant, $\epsilon^{ikl}$ is the Levi-Civita symbol, $\rho^j$ is the vector mass density and $J^{ij}$ is the mass current density. The symbol $(\cdots)$ denotes symmetrization of the first and last indices, i.e., $i$ and $j$.

To construct a lagrangian formulation that describes the GEM equations, fields ${\cal E}^{ij}$ and ${\cal B}^{ij}$ are defined as (details are given in \cite{Khanna})
\bea
{\cal E}&=&-\mathrm{grad}\,\varphi-\frac{\partial \tilde{\cal A}}{\partial t},\\
{\cal B}&=&\mathrm{curl}\,\tilde{\cal A},
\eea
where $\tilde{\cal A}$ is a symmetric rank-2 tensor field, gravitoelectromagnetic tensor potential, with components ${\cal A}^{\mu\nu}$  and $\varphi$ is the GEM vector counterpart of the electromagnetic scalar potential $\phi$. The ${\cal E}_{ij}$ and ${\cal B}_{ij}$ tensor fields are elements of a rank-3 tensor, the gravitoelectromagnetic tensor, $F^{\mu\nu\alpha}$, defined as
\bea
F^{\mu\nu\alpha}=\partial^\mu{\cal A}^{\nu\alpha}-\partial^\nu{\cal A}^{\mu\alpha},
\eea
where $\mu, \nu,\alpha=0, 1, 2, 3$. The non-zero components of ${F}^{\mu\nu\alpha}$ are ${F}^{0ij}={\cal E}^{ij}$ and ${F}^{ijk}=\epsilon^{ijl}{\cal B}^{lk}$ where $i, j=1, 2, 3$. With these definitions the Maxwell-like equations, given in Eqs. (\ref{01})-(\ref{04}), are written in a covariant form as
\bea
\partial_\mu{F}^{\mu\nu\alpha}&=&4\pi G{\cal J}^{\nu\alpha},\\
\partial_\mu{\cal G}^{\mu\langle\nu\alpha\rangle}&=&0,
\eea
where ${\cal J}^{\nu\alpha}$ depends on the mass density, $\rho^i$, and the current density $J^{ij}$.  Here ${\cal G}^{\mu\nu\alpha}$ is the dual GEM tensor, that is defined as
\bea
{\cal G}^{\mu\nu\alpha}=\frac{1}{2}\epsilon^{\mu\nu\gamma\sigma}\eta^{\alpha\beta}{F}_{\gamma\sigma\beta}.
\eea

The GEM lagrangian is constructed considering the symmetric gravitoelectromagnetic tensor potential $A_{\mu\nu}$ as the fundamental field that describes the gravitational interaction. Then
\bea
{\cal L}_G=-\frac{1}{16\pi}{F}_{\mu\nu\alpha}{F}^{\mu\nu\alpha}-G\,{\cal J}^{\nu\alpha}{\cal A}_{\nu\alpha}.\label{L_G}
\eea
In the weak field approximation, $A_{\mu\nu}$ has similar symmetry properties to those of $h_{\mu\nu}$, which is a tensor defined in Einstein Gravity. However our approach is different, since the nature of $A_{\mu\nu}$ is different from $h_{\mu\nu}$. In addition, the tensor potential is connected directly with the description of the gravitational field in flat spacetime and it has nothing to do with the perturbation of the spacetime metric.

\section{Lorentz-violating GEM}

The lagrangian that describes the graviton-fermion interaction is 
\bea
{\cal L}&=&-\frac{1}{16\pi}F_{\mu\nu\alpha}F^{\mu\nu\alpha}-\bar{\psi}\left(i\gamma^\mu \overleftrightarrow{D_\mu}-m\right)\psi,\label{eq1}
\eea
where $\psi$ is the fermion field with $\bar{\psi}=\psi^\dagger \gamma_0$, $m$ is the fermions mass, $\gamma^\mu$ are Dirac matrices and $D_\mu$ is the covariant derivative. Our objective is to calculate the Lorentz violation effects in the graviton-fermions interaction.  Besides the investigations into the structure of the SME, where all interactions of the graviton coupled to the SM fields are established \cite{Kos0}, there is other propose to examine Lorentz-violating developments out of this broad framework. This alternative involves non-minimal coupling terms that modify the vertex interaction between fermions and gravitons.
To introduce the Lorentz-violating term in the lagrangian (\ref{eq1}) the usual covariant derivative will be modified by a non-minimal coupling term, i.e.,
\bea
\overleftrightarrow{D_\mu}=\overleftrightarrow{\partial_\mu}-\frac{1}{2}igA_{\mu\nu}\overleftrightarrow{\partial^\nu}+\frac{1}{4}\bigl(k^{(5)}\bigl)_{\mu\nu\alpha\beta\rho}\gamma^\nu F^{\alpha\beta\rho},\label{der}
\eea
where $g=\sqrt{8\pi G}$ is the gravitational coupling constant and $\bigl(k^{(5)}\bigl)_{\mu\nu\alpha\beta\rho}$ is a tensor that belongs to the gravity sector of the non-minimal SME with mass dimension $d=5$ \cite{QG-K}. There is an analogous Lorentz-violating term in the electromagnetic sector of the non-minimal SME. Then GEM with Lorentz violation is part of the SME. Using eq. (\ref{der}) the lagrangian becomes
\bea
{\cal L}&=&-\frac{1}{16\pi}F_{\mu\nu\alpha}F^{\mu\nu\alpha}-\frac{i}{2}\left(\bar{\psi}\gamma^\mu \partial_\mu\psi-\partial_\mu\bar{\psi}\gamma^\mu\psi \right)+m\bar{\psi}\psi\nonumber\\
&-&\frac{g}{4}A_{\mu\nu}\left(\bar{\psi}\gamma^\mu\partial^\nu\psi-\partial^\mu\bar{\psi}\gamma^\nu\psi\right)-\frac{1}{4}\bigl(k^{(5)}\bigl)_{\mu\nu\alpha\beta\rho} F^{\alpha\beta\lambda}\bar{\psi}\sigma^{\mu\nu}\psi,
\eea 
where $\sigma^{\mu\nu}=\frac{i}{2}\left(\gamma^\mu\gamma^\nu-\gamma^\nu\gamma^\mu\right)$ and the definition $A\overleftrightarrow{\partial^\mu}B\equiv\frac{1}{2}\left(A\partial^\mu B-\partial^\mu AB\right)$ is used. Then the interaction part of the lagrangian is
\bea
{\cal L}_I=-\frac{g}{4}A_{\mu\nu}\left(\bar{\psi}\gamma^\mu\partial^\nu\psi-\partial^\mu\bar{\psi}\gamma^\nu\psi\right)-\frac{1}{4}\bigl(k^{(5)}\bigl)_{\mu\nu\alpha\beta\rho} F^{\alpha\beta\lambda}\bar{\psi}\sigma^{\mu\nu}\psi.
\eea
The first term describes the usual interaction between gravitons and fermions and the second term is a new interaction produced by the non-minimal coupling. This new interaction describes the non-minimal coupling between the GEM field and the fermion bilinear. It is similar to the non-minimal coupling between the electromagnetic field and the fermion bilinear \cite{Kost_H}. Then the Lorentz-violating GEM coefficient $\bigl(k^{(5)}\bigl)_{\mu\nu\alpha\beta\rho}$ is analogous to the electromagnetic coefficient $H_F^{(5)\mu\nu\alpha\beta}$. It is one more confirmation that GEM is a gravity theory analogous to the electromagnetic theory.

From the interaction lagrangian the vertices are 
\bea
\bullet&\rightarrow& -\frac{ig}{4}\left(\gamma^\beta p_1^\rho+p_2^\beta\gamma^\rho\right)\\
\circ&\rightarrow& -\frac{1}{2}\bigl(k^{(5)}\bigl)^{\mu\nu\alpha\beta\rho}\sigma_{\mu\nu} q_\alpha.
\eea

Our object is to calculate the differential cross section for the gravitational Bhabha scattering with Lorentz violation.  Here the momentum transfer, $q_\alpha$ is considered as $q_\alpha=(\sqrt{s},0)$, with $s$ being the energy in the center of mass.

\section{Differential cross section}

The differential cross section for GEM Bhabha scattering is calculated. Our objective is to analyze how the new vertex modifies the scattering process. This process is represented in FIG. 1.
\begin{figure}[h]
\includegraphics[scale=0.3]{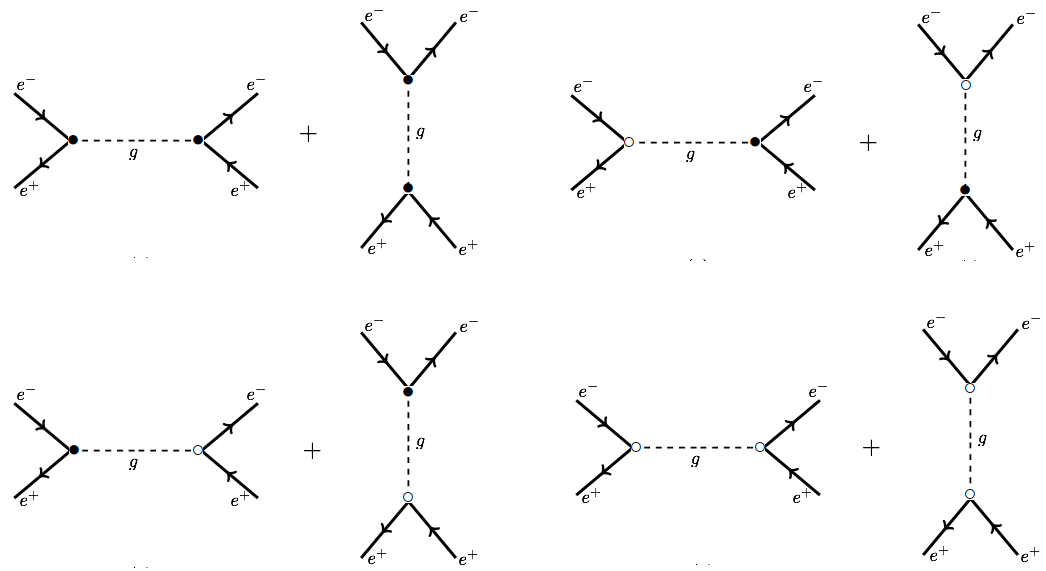}
\caption{GEM Bhabha Scattering with one graviton exchange }
\end{figure}

The graviton propagator is given by
\bea
D_{\mu\nu\alpha\rho}=\frac{i}{2q^2}\left(g_{\mu\alpha}g_{\nu\rho}+g_{\mu\rho}g_{\nu\alpha}-g_{\mu\nu}g_{\alpha\rho}\right)
\eea
and for the vertex following notation is used
\bea
V^{\beta\rho}_{(0)}&=&-\frac{ig}{4}\left(\gamma^\beta p_1^\rho+p_2^\beta\gamma^\rho\right)\\
V^{\beta\rho}_{(1)}&=& -\frac{1}{2}\bigl(k^{(5)}\bigl)^{\mu\nu\alpha\beta\rho}\sigma_{\mu\nu} q_\alpha.
\eea

The calculation is carried out in the center of mass frame (CM), where the particles are labeled $e^-(p_1)+e^+(p_2)\rightarrow e^-(q_1)+e^+(q_2)$. In the CM frame we have
\bea
p_1&=&(E,\vec{p}),\quad\quad p_2=(E,-\vec{p}),\nonumber\\
q_1&=&(E,\vec{p'})\quad\quad\mathrm{and}\quad\quad q_2=(E,-\vec{p'}),
\eea
where $|\vec{p}|^2=|\vec{p'}|^2=E^2$, $\vec{p}\cdot\vec{p'}=E^2\cos\theta$ and
\bea
p_1\cdot p_2=q_1\cdot q_2&=&2E^2, \quad\quad p_1\cdot q_1=E ^2(1-cos\theta)\nonumber\\
p_1\cdot q_2=q_1\cdot p_2&=&2E^2, \quad\quad p_2\cdot q_2=E ^2(1-cos\theta).
\eea

The differential cross section is defined as
\bea
\frac{d\sigma}{d\Omega}=\left(\frac{\hbar^2 c^2}{64\pi^2 s}\right)\cdot\frac{1}{4}\sum_{spins}\bigl|{\cal M}\bigl|^2,\label{cs}
\eea
where an average over the spin of the incoming particles and summing over the outgoing particles is considered. The transition matrix element is written using Feynman diagrams (FIG. 1). The matrix element of Bhabha scattering is given as
\bea
{\cal M}={\cal M}_{ann}+{\cal M}_{exc},
\eea
where ${\cal M}_{ann}$ and ${\cal M}_{exc}$ are the transition amplitudes due to the annihilation process and the exchange process, respectively. From Feynman diagrams these transition amplitudes are
\bea
{\cal M}_{ann}&=&\sum_{a,b}\left[\bar{v}(p_2)V^{\mu\nu}_{(a)}u(p_1)\right]D_{\mu\nu\alpha\rho}(q)\left[\bar{u}(q_1)V^{\alpha\rho}_{(b)}v(q_2)\right],\\
{\cal M}_{exc}&=&-\sum_{a,b}\left[\bar{u}(q_1)V^{\alpha\rho}_{(a)}u(p_1)\right]D_{\mu\nu\alpha\rho}(q)\left[\bar{v}(p_2)V^{\mu\nu}_{(b)}v(q_2)\right],
\eea
with $a,b=0, 1$. Here $u$ and $v$ are spinors for the electron and the positron, respectively. For evaluating the differential cross section the relevant quantity is $|{\cal M}|^2=\sum{\cal M}{\cal M}^*$, where the sum is over spins. The calculation is carried out using the completeness relation:
\bea
\sum_{s_1} u(p_1, s_1)\bar{u}(p_1, s_1)&=&\slashed{p}_1+m \quad\quad \mathrm{and}\nonumber\\
\sum_{s_1} v(p_1, s_1)\bar{v}(p_1, s_1)&=&\slashed{p}_1-m.
\eea
And using the relation
\bea
\bar{v}(p_2)\gamma_\alpha p_{1\rho}u(p_1)\bar{u}(p_1)\gamma^\alpha p_1^\rho v(p_2)=\mathrm{tr}\left[\gamma_\alpha p_{1\rho}u(p_1)\bar{u}(p_1)\gamma^\alpha p_1^\rho v(p_2)\bar{v}(p_2)\right]
\eea
the summation is accomplished. Henceforth the electron mass is ignored since all the momenta are large compared with the electron mass.

In order to present the differential cross section, concerning the beam orientation in relation to the background tensor, consider that the beam is perpendicular to the background, i.e., $\bigl(k^{(5)}\bigl)_{\mu\nu\alpha\beta\rho}\,p^\rho=0$. The differential cross section to second order is
\bea
\left(\frac{d\sigma}{d\Omega}\right)&=&\left(\frac{d\sigma}{d\Omega}\right)_{GEM}\Biggl[1+\frac{\bigl(k^{(5)}\bigl)^2}{g^2{\cal O}}\Bigl(128\cos^2\theta\sin^6(\theta/2) -4924\cos\theta + 1552 \cos 2\theta\nonumber\\ 
&-& 630 \cos 3\theta +140\cos 4\theta +14\cos 5\theta+ 3876\Bigl)\Biggl],\label{odd1}
\eea
where ${\cal O}\equiv 1864\cos\theta+540\cos 2\theta+56\cos 3\theta+5\cos 4\theta+1631$. Here  $\left(\frac{d\sigma}{d\Omega}\right)_{GEM}$ is the differential cross section for the GEM field \cite{AFS}, Lorentz invariant case, and is given by
\bea
\left(\frac{d\sigma}{d\Omega}\right)_{GEM}=-\frac{g^4E^4}{n\,\pi^2\,s}\frac{\left(1864\cos\theta+540\cos 2\theta+56\cos 3\theta+5\cos 4\theta+1631\right)}{(\cos\theta-1)^2},
\eea
with $n$ being a numerical factor defined as $n\approx6,6\times 10^4$. Here $\hbar=c=1$ is used. Then the first term in eq. (\ref{odd1}) is the usual GEM differential cross section at the lowest order and the second term gives the contribution of the Lorentz-violating part.

\section{Conclusions}

 Fundamental theories in particle physics have concentrated on experiments at energies that are available from accelerators. These have provided a great deal of data that is understood by Lorentz covariant standard model theories of particle physics. However lately it has been surmised that it will be interesting to consider the possibility of there being Lorentz symmetry violation at high energies, i.e. at Planck scale. This gave us so called the Standard Model Extension (SME), that is the standard model with some broken symmetries. In principle an extensive array of operators that break symmetry can be written down.  Initial attempts considered such extensions only for particle physics. Later these have been considered for the Einstein theory of Gravity. The Gravitoelectromagnetism (GEM), that has similarities to the Maxwell equations for Electromagnetic field, is an approximation of the general relativity that is valid only in a suitable situation, such as weak-field approximation or if degrees of freedom are dependent on the Weyl tensor describing tidal forces. In this paper the Lagrangian formulation of GEM is used to study Bhabha scattering with the Lorentz violating components developed for the standard model extension (SME). The Lorentz violation is introduced via the non-minimal coupling with covariant operators. This new coupling implies the insertion of a new vertex, thus increasing the number of Feynman diagrams representing the process. The contributions of the non-minimal Lorentz-violating terms on the differential cross section are calculated for the gravitational Bhabha scattering. GEM allows electromagnetic scattering processes with gravitons as intermediate state in addition to the photon electromagnetic scattering. Our results show that the Lorentz violation contribution for gravitational Bhabha scattering is similar to the corrections in the electromagnetic case. This is one more result that gives emphasis to the analogy between these two theories. Although the present study is entirely theoretical, it is possible that in  future this result may be used to test GR, since GEM is an approximation of GR. In addition, constraints on the Lorentz-violating parameter of the model may be obtained if the precision of the measurements will improve significantly. Here the cross section is calculated in the CM frame. However the coefficients in the CM frame are not constant because all experiments with beams involve non-inertial laboratories on the Earth, which is rotating in the standard Sun-centered inertial frame (SCF). Then CM coefficients must be converted to SCF coefficients as discussed in \cite{Kost_H, Kost2002, Kost1998}.

\section*{Acknowledgments}

It is a pleasure to thank V. A. Kosteleck\'y for useful remarks about the Lorentz-violating coefficient for GEM field.

\end{document}